# Compatible and Usable Mandatory Access Control for Good-enough OS Security

Zhiyong Shan
Computer Science Department, Renmin University of China
Computer Science Department, State University of New York at Stony Brook
Beijing, China
Email: zyshan2003@hotmail.com

*Abstract*—OS compromise is one of the most serious computer security problems today, but still not being resolved. Although people proposed different kinds of methods, they could not be accepted by most users who are non-expert due to the lack of compatibility and usability. In this paper, we introduce a kind of new mandatory access control model, named CUMAC, that aims to achieve good-enough security, high compatibility and usability. It has two novel features. One is access control based on tracing potential intrusion that can reduce false negatives and facilitate security configuration, in order to improve both compatibility and usability; the other is automatically figuring out all of the compatibility exceptions that usually incurs incompatible problems. The experiments performed on the prototype show that CUMAC can defense attacks from network, mobile disk and local untrustable users while keeping good compatibility and usability.

## I. INTRODUCTION

In 2004 the USA Today reported that hackers were selling computing time on large networks of compromised machines, which are believed to be used for activities such as online extortion, delivering unsolicited email advertisements, and identity theft through fraudulent web sites. Some valuable security countermeasures, such as antivirus software, network Firewalls, and antispyware software, may deal with this problem. However, the F.B.I. conducted a survey of U.S. businesses, which found that 87% of the companies surveyed experienced at least one computer security compromise in 2005, and most of these companies use those security countermeasures to prevent compromises. This evidence shows that the current state-of-the-art in computer security defenses does not prevent all security compromises.

To solve the problem, Mandatory Access Control (MAC) is a good choice, which can restrict privileged processes so that damage resulting from compromise is limited. Over the last 30 years, many projects have demonstrated useful MAC features on operating systems. Recent examples include DTE [1], LOMAC [2], Security-Enhanced Linux [3] and UMIP [4]. However, despite their success, these demonstrations have some problems hesitating non-expert users to accept them. For example, the policy interface of SELinux is daunting even for security experts [4]; the recent UMIP demonstrating a usable mandatory integrity protection in OS kernel still suffers the problem of "self revocation"[2] that downgrades its compatibility.

This paper presents our novel form of access control, Compatible and Usable Mandatory Access Control (CUMAC), which aims to prevent system compromise in a both compatible and usable manner so as to tackles the OS compromise problem. It is designed to meet three specific goals. First, good-enough security, i.e. CUMAC can prevent remote or local attackers from taking over a host, but is not required to reach a higher level of security than necessary as well. Second, high compatibility, i.e. CUMAC must be compatible with the existing deployed Commercial Off-The-Shelf (COTS) software. This goal implies that CUMAC should not require the replacement or modification of existing software and its configurations in order to operate smoothly, as well as not cause failures in the working software. Third, high usability, i.e. CUMAC must be largely transparent to the user. This goal implies that the user should not be required to learn new behaviors in order to work in a CUMAC-enhanced environment, and configuring a CUMAC should be an automatic progress.

The basic CUMAC policy is as follows: Each process and file has two states, which are either potential intrusion or non-intrusion. From each potential intrusion entrance in OS, CUMAC labels the relevant processes and executable-files as potential intrusion. Tracing the activities of these processes or processes derived from the potential intrusion files, CUMAC labels the processes they forked and communicated with as potential intrusion, as well as labels the executable-files they created or modified as potential intrusion. When potential intrusion processes request to launch security critical operations, CUMAC refuses them so as to prevent possible host compromise.

Comparing with previous MAC systems, CUMAC has two novel features. One is access control based on tracing potential intrusion. As most of the traditional MAC systems require manually configuring entity labels, the biggest advantage of tracing potential intrusion is to label OS entities automatically. Meanwhile, access control based on tracing potential intrusion can reduces false negatives that are the root cause of incompatibility in a MAC enforced system. The other novelty is to automatically figure out all compatibility exceptions usually incurring incompatibility problems by a carefully designed mechanism. For most of the MAC policies, compatibility exception is inevitable and difficult to be found out. However, so far rare research has focused on this issue.

This research is supported by the Natural Science Foundation of China under grant number 60703103/60873213, the 863 High Technology Foundation of China under grant number 2007AA01Z414, and the Research Foundation of Renmin University of China under grant number 06XNB053

Especially, we do not find any research on the automated compatibility exception handling mechanism.

The rest of the paper is organized as follows. Section 2 presents our considerations before construct the model. Section 3 describes the basic CUMAC model and exception mechanism. Section 4 discusses its prototype and tests on Linux. We present related works and make a conclude in Section 5.

## II. CONSIDERATIONS

Before presenting the model, the goals of the model should be analyzed deeply in order to find out proper ways to achieve them.

### A. Good-enough Security

Sandhu [1] observed that "cumbersome technology will be deployed and operated incorrectly and insecurely, or perhaps not at all." and suggested the following adaptation for the information security business: "Everything should be made as secure as necessary, but not securer." This is the essence of good-enough information security. In the other words, good-enough security requires that security in a manner that tradeoff security for other competing concerns.

From the perspective of OS, good-enough security means necessary security along with good compatibility and usability. Since the most serious security problem in OS is compromise, the necessary security in OS would be prevention of it. As hundreds of thousands of computers managed by users without any knowledge in system security, one thus needs a security protection system with a high level of usability. On the other hand, no compatibility no usability, usability is based on compatibility. For instance, there is no sense to discuss the usability of a web server if it even cannot run smoothly. So, the protection mechanism should tradeoff security for both compatibility and usability while compatibility is prior to usability.

Therefore, our objective is to build a system compatible with existing COTS software and easy to be configured, while greatly increases security level by narrowing the attack channels. However, we do not aim to provide a protection system with theoretically very strong security guarantees, whereas at the same time it is incompatible with existing applications and requires huge effort to configure it correctly.

### B. Compatibility

Traditional MAC models, e.g. BLP, BIBA, DTE, can offer high security to OS through proper configuration. On the other hand, they also bring incompatibility into OS, causing many application failures [1] [2]. As a result, there are limitations when applying them to build a system with both high security and compatibility.

From the perspective of intrusion detection, the incompatibility of traditional MAC comes from false negative, i.e. refusing the accesses that should be allowed, which leads to disturbing applications' running on the OS. For example, the self-revocation problem [2] in Low-Water-Mark model is a typical false negative, refusing a process to write a file created by the process if it has read a lower level file before writing. From the perspective of intrusion detection, the write operation should not be refused if the lower level file actually does not contain harmful data. Another example of typical false negative is the shared /tmp directory [1] on a BLP enforced OS. According to BLP model, a process with different sensitive level or category can not write or read the entire system shared /tmp directory. From the perspective of intrusion detection, the process does not necessarily represent an intruder so that the read or write operation should not be simply refused. Since there are much more similar shared entities on OS such as shared files, devices, pipes and memory, these shared entities will make lots of software abnormally fail.

However, the false negative is an inherent defect in the traditional MAC models. Because, traditional MAC makes access decision based on the labels of subject and object that are usually static or limitedly dynamic. As a result, MAC cannot change the labels dynamically to reflect intrusion progress in OS, i.e. identify intrusion entities in OS. In other words, MAC cannot correctly recognize intrusion entities and non-intrusion entities at the time to make access decision. Therefore, MAC often fails to make correct access decisions in order to avoid false negatives.

Thereby, we know how to improve the compatibility of an access control model. That is to dynamically change the subject and object labels according to the intrusion spread progress in OS.

### C. Usability

On other side, traditional MAC also can not provide good usability to general users due to requiring complicated configuration and usage method different from convention. In modern OS, as there are wide kinds of entities including user, process, file, directory, device, pipe, signal, shared memory and socket, etc, it is difficult for a general user to correctly configure labels for all entities without leaving security vulnerabilities. Meanwhile, after enforcing MAC, users must break their usage convention and learn how to use MAC. Consequently, the ideal way for access control model to provide good usability is automatically initializing and changing labels of entities and saving them through conventional security mechanisms user familiar with.

From the perspective of defeating intrusion, only two types of labels, i.e. intrusion and non-intrusion, are necessary. Thus, automatic label initialization and change can be achieved by tracing and identifying intrusion entities.

## III. CUMAC MODEL

Though the description of CUMAC model is based on the design for Linux/Unix OS, we believe that the model is also suitable for non-Unix operating systems such as the Microsoft Windows family with specific changes. Investigating the applicability of CUMAC for Microsoft Windows is beyond the scope of this paper.

### A. Intrusion Analysis

According to the analysis above, improving both capability and usability of an access control model needs to trace intrusions in OS so as to automatically set security label for OS entities, i.e. identifying intrusion entities. Usually, intrusions on OS consist of three phases: intrusion entrance, intrusion spread and security critical operation.

Intrusion entrance represents an entry point through which intruder gets access to OS, for example, communicating with remote host. Conservatively, any session on OS steming from such type of entrances should be considered as potential intrusion. In other words, the processes belonging to such session and files modified by these processes also should be considered as potential intrusion.

Security critical operation represents operations through which intruder can illegally compromise the system and read sensitive information, including system privileged operations, writing system important executable and configuration files, as well as reading files containing secrecies. In most cases, security critical operation is the target or necessary step of an intrusion. Without successfully performing the planned security critical operation, an intrusion can not achieve its goals. Consequently, access control mechanism must prevent any potential intrusion process from executing security critical operation in order to protect whole system.

Intrusion spread represents intrusion steps from intrusion entrance to security critical operation. These steps can be uncovered by the method analyzing OS level information flow, which is proposed and verified by recent years' research projects [8][9][10].

*B. Model Description*

Corresponding to the three phases of intrusion on OS, the CUMAC model is comprised of intrusion entrance, intrusion spread and security critical operations, as well as protection rules used to prevent potential intrusion process from executing security critical operations. Figure 1 shows the basic phases of the CUMAC model.

Firstly, CUMAC traces potential intrusion from three types of possible intrusion entrances, involving communicating with remote host, mounting mobile storage and untrustable user login. Obviously, communicating with remote host is the most frequently exploited way by intruders. Meanwhile, mounting mobile storage also can introduce malware into system due to the mobile storage might contain malware from other compromised machine. Finally, local untrustable user can upgrade him as super user through specific local attack. Consequently, CUMAC treat following three entities as the start points for tracing potential intrusion:

- Processes conducted remote communication;

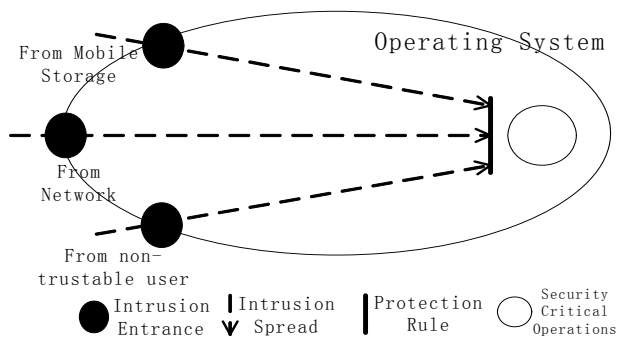

Figure 1. CUMAC Model

- Executable files copied from a mounted mobile storage;
- Processes coming from the login of an untrustable user.

Secondly, CUMAC traces intrusion spreading activities in OS by employing OS level information flow analysis method [8] [9] [10]. It flags following entities as potential intrusion:

- Processes spawned by a potential intrusion process;
- Executable files created or modified by a potential intrusion process;
- Processes communicated with a potential intrusion process;
- Processes launched from a potential intrusion file.

Lastly, CUMAC blocks potential intrusion processes to perform security critical operations. There are three categories of security critical operations including:

- System privileged operations that only can be executed by super user;
- Write files requiring integrity protection, such as executable files and important configuration files;
- Read files requiring sensitivity protection, such as files containing system secrecy or user's secrecy.

To improve usability, CUMAC uses existing security mechanism to recognize security critical operations. For Linux platform, CUMAC treats process capabilities as system privileged operations, treats files without "write" permission for other users as files requiring integrity protection, and treats files without "read" permission for other users as files requiring sensitivity protection.

In summary, the basic CUMAC model works in this way: Each process and file has two states, which are either potential intrusion or non-intrusion. From each potential intrusion entrance in OS, CUMAC labels the relevant processes and executable-files as potential intrusion. Tracing the activities of these processes or processes derived from the potential intrusion files, CUMAC labels the processes they forked and communicated with as potential intrusion, as well as labels the executable-files they created or modified as potential intrusion. When potential intrusion processes request to launch security critical operations, CUMAC refuses them so as to prevent possible host compromise.

*C. Exception Mechanism*

With the CUMAC model presented above, computer compromise problem can be resolved. But yet there are some special accesses causing false negative, called exceptional access. Exceptional access is not intrusion but will be refused by access control mechanism and cause applications fail. For example, when super user tries to upgrade the system from remote machine, CUMAC will regard it as potential intrusion and then refuse him to over write executable files. Although our experiments showed that exceptional accesses amount to very little, an exception mechanism to deal with them is still appealed since CUMAC aims to achieve high compatibility and usability. Moreover, papers [4] [1] also pointed out the importance of exception mechanism. Nevertheless, so far there

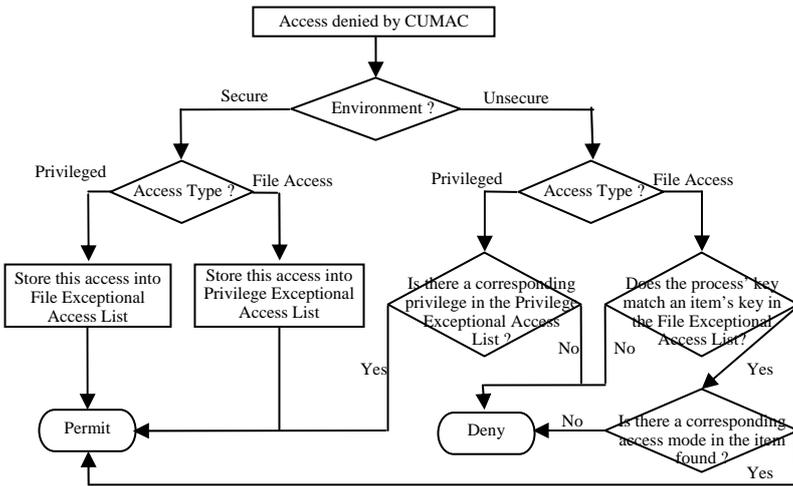

Figure 2. Exception Mechanism Working Flow

is rare research work focused on this topic, especially automatic exception mechanism.

Usually, exception mechanism can be built as a type of white-list on which all exceptional accesses are recorded. But it is not feasible to be enforced on a system with many kinds of software installed, because user could not find out all exceptional accesses. Therefore, the ideal method is to automatically figure out all exceptional accesses.

Concretely, we imagine the exception mechanism works in this way: in a secure environment that is sure no intrusion, exception mechanism automatically records and stores all the exceptional accesses that is refused by the access control mechanism into an exceptional access list; then, in an unsecure environment that might have intrusion, exception mechanism authorizes exceptional accesses according to the exceptional access list.

In order to build up the exception mechanism, we performed some experiments from which two realities are uncovered. One is that each exceptional access is necessarily generated by certain application. In other words, every exceptional access belongs to a specific application. The other is that all exceptional accesses can be divided into two categories: privileged operation related exceptional access and file operation related exceptional access.

Based on the analysis and experiment results above, we designed the exception mechanism. It has four basic elements as follows:

(1) Key. Every application has a unique key to identify itself, which comes from the file node number of the application's executable file. Any process started from an executable file will automatically inherit the key. Exception mechanism authorizes an exceptional access according to process' key.

(2) File Exceptional Access List. Every file or directory has such a list and stores it in the file node as a security property. Every item of the list is a pair containing a key and a vector recording permitted access modes to the file or directory. In other words, the item actually describes which application can perform what kinds of accesses to the file or directory.

(3) Privilege Exceptional Access List. Every application has such a list and stores it in executable file's file node as security property. Every item of the list represents a privilege the application owns. Processes of the application will inherit the privileges from its executable file.

(4) Environment Bit. There is only one such a bit for the entire system to mark the running environment of the system. As presented above, in a secure environment, exception mechanism records exceptional accesses into the two exceptional access lists described above; in an unsecure environment, exception mechanism authorizes exceptional access according to the two lists. Usually, the bit is set by system administrator.

The work flow of the exception mechanism is depicted in Figure 2. When the system is running with the bit value representing secure environment, all the accesses that should be refused by access control mechanism are permitted and recorded into the two lists as exceptional accesses. When the system is running with the bit value representing unsecure environment, for each accesses refused by access control mechanism, exception mechanism will further check if it is an exceptional access recorded in the two lists, and permit it if it is an exceptional access.

IV. PROTOTYPE

To verify the capability of CUMAC model to effectively prevent host compromise while keeping high compatibility and usability, we implemented a CUMAC prototype in Linux kernel 2.6.9 through the means of Linux Security Module (LSM).

A. Implementation

The CUMAC implementation consists of three parts: decision, exception and information. The decision part handles access requests intercepted from system calls. When making a decision, it firstly reads the CUMAC attributes of subject and object from information part, and then decides whether to permit the access and whether to modify the CUMAC attributes according to the CUMAC model rules presented at section 3.2. In the case of denying the access, the decision part will try to forward the access request to the exception part to check if the access is an exception. The information part maintains CUMAC attributes in memory, containing key, potential intrusion flags and exceptional access lists. The whole implementation is encapsulated in a LSM, and does not impose any modifications on the Linux kernel code, thus it is highly compatible with Linux kernel.

B. Evaluation

In order to experimentally evaluate the CUMAC model, we established two servers configured with Fedora Core 3 whose kernel version is 2.6.9, and loaded our LSM module during system boot. We tested the CUMAC prototype from four dimensions: compatibility, usability, security, and performance.

High compatibility requires that existing deployed COTS software can run on the CUMAC prototype without significant incompatibility problems. On the two servers deployed, we installed many commonly used network applications and local applications e.g. apache, mysql, samba, ftp, telnet, mozilla, dhcp, sendmail, KDE desktop, open office, webmin, ssh, yum, etc. Then, we used them to send mails, go on Internet, install software remotely, share files remotely, manage user accounts remotely and move file by USB disk. The system works well for the last four months, without modifications of existing software and running failures.

Usability need to be evaluated from two aspects including the amount of configuration work and the impaction on user's convention. The configuration work of CUMAC is only to set the environment bit, and all other CUMAC information is automatically configured. The running and configuration of CUMAC is transparent to users who thus only need to manage traditional DAC mechanism.

To verify the functionality of security protection, we carried out three experiments including defeating intrusion from USB disk, local untrustable user and network. First, we run a rootkit named adore-ng-2.6 from a mounted USB disk, trying to install a kernel module. The CUMAC module then refuses the insmod request that usually is a critical operation to control the whole system. Second, we login the system with a non-administration account and invoke ptrace to exploit a race condition so as to cause the target kernel to hang. The CUMAC successfully refuses the user to execute ptrace function. Third, we download a rootkit named knark [11] through Mozilla web browser and try to run it. CUMAC then successfully blocks its operation to install a kernel module.

In addition, CUMAC has little impact on Linux performance. First, the decision are simple as they only compare flags of the subject and object; second, getting decision data is also fast in that potential intrusion flags are stored in the security field of the concerned kernel data structures. In the performance test, we compared the performance of two kernels: the original Linux kernel and the CUMAC-enforced kernel. This test uses "kernel compile" [29] as test method, a broadly accepted method for testing the general performance of Linux. The result shows that CUMAC-enforced Linux kernel only imposes 2.1% performance overhead comparing to the original Linux kernel.

Based on the above tests, we can safely say that the CUMAC-enforced Linux can block system compromise operations while keeping high compatibility and usability, as well as incur trifle performance overhead.

V. RELATED WORK AND SUMMARY

There are two most related works, i.e. UMIP and LOMAC. UMIP evolves from BIBA model while aims to add usable mandatory integrity access control to operating system. But it suffers several shortcomings. First, it can not address the self-revocation problem that is an inherent defect in BIBA model so that still has incompatibility problem. Second, it requires user to manually figure out all exceptional accesses, thus fail to offer perfect usability to general user. Third, it confuses low integrity level with potential intrusion so that fails to trace intrusion when low integrity level processes write low integrity level files. Lastly, it only can defense attack from network.

Similar to UMIP, LOMAC also try to introduce a type of mandatory integrity protection mechanism in OS while compatible with widespread commercial software. But it also suffers self-revocation problem and can not automatically configure security labels.

Comparing with these two projects, CUMAC is a type of novel access control, providing high compatibility, usability and good enough security protection to user. It employs two novel methods. One is tracing intrusion to automatically flag potential intrusion entities in order to prevent potential intrusion processes from executing security critical operations. The other is automatically figuring out exceptional accesses so as to eliminate compatibility exceptions. In short, CUMAC is a new access control model performing access control based on tracing potential intrusion. Test results show that CUMAC can defeat attacks from mobile disk, untrustable user and network by refusing their requests to perform security critical operation that results in whole system compromise, while offer high compatibility to COTS software and high usability to ordinary users.

REFERENCES

[1] Lee Badger, Daniel F. Sterne, David L. Sherman, Kenneth M. Walker, and Sheila A. Haghighat. A domain and type enforcement UNIX prototype. In Proc. of the 5th USENIX UNIX Security Symposium, June 1995.
[2] Timothy Fraser. LOMAC: Low Water-Mark Integrity Protection for COTS Environments. In Proceedings of the IEEE Symposium on Security and Privacy, Oakland, CA, May 2000.
[3] P. A. Loscocco and S. D. Smalley. Integrating Flexible Support for Security Policies into the Linux Operating System. In Proceedings of the FREENIX Track: USENIX Annual Technical Conference, June 2001.
[4] Ninghui Li, Ziqing Mao, Hong Chen, "Usable Mandatory Integrity Protection for Operating Systems," sp, pp. 164-178, 2007 IEEE Symposium on Security and Privacy (SP '07), 2007
[5] R. Sandhu. Good-enough security: Toward a pragmatic business-driven discipline. IEEE Internet Computing, 7(1):66–68, Jan. 2003.
[6] V.D. Gligor, E.L. Burch, G.S. Chandersekaran, R.S. Chapman, L.J. Dotterer, M.S. Hecht, W.D. Jiang, G.L. Luckenbaugh, N. Vasudevan, "On the Design and the Implementation of Secure Xenix Workstations," sp, p. 102, 1986 IEEE Symposium on Security and Privacy, 1986
[7] Timothy Fraser. LOMAC:MAC You Can LiveWith. In Proceedings of the FREENIX Track, USENIX Annual Technical Conference, Boston, MA, June 2001
[8] ST King, PM Chen. Backtracking intrusions. ACM Transactions on Computer Systems (TOCS), 2005.
[9] ZHU, N. AND CHIUEH, T. 2003. Design, implementation, and evaluation of repairable file service. In Proceedings of the 2003 International Conference on Dependable Systems and Networks (DSN). 217–226.
[10] Kamran Farhadi Zheng Li Ashvin Goel, Kenneth Po and Eyalde Lara, "The taser intrusion recovery system," in Proceedings of the twentieth ACM symposium on Operating systems principles, 2005.
[11] Plaguez. Weakening the linux kernel. Phrack, 8(52), January 1998
[12] Zhiyong Shan, Tanzirul Azim, Iulian Neamtiu. Finding Resume and Restart Errors in Android Applications. ACM Conference on Object-Oriented Programming, Systems, Languages & Applications (OOPSLA '16), November 2016. Accepted.
[13] Zhiyong Shan, I. Neamtiu, Z. Qian and D. Torrieri, "Proactive restart as cyber maneuver for Android", Military Communications Conference, MILCOM 2015 - 2015 IEEE, Tampa, FL, 2015, pp. 19-24.
[14] Jin, Xinxin, Soyeon Park, Tianwei Sheng, Rishan Chen, Zhiyong Shan, and Yuanyuan Zhou. "FTXen: Making hypervisor resilient to hardware faults on relaxed cores." In 2015 IEEE 21st International Symposium on High Performance Computer Architecture (HPCA'15), pp. 451-462. IEEE, 2015.
[15] Zhiyong Shan, Xin Wang, Tzi-cker Chiueh: Shuttle: Facilitating Inter-Application Interactions for OS-Level Virtualization. IEEE Trans. Computers 63(5): 1220-1233 (2014)
[16] Zhiyong Shan, Xin Wang. Growing Grapes in Your Computer to Defend Against Malware. IEEE Transactions on Information Forensics and Security 9(2): 196-207 (2014)
[17] Zhiyong Shan, Xin Wang, Tzi-cker Chiueh: Malware Clearance for Secure Commitment of OS-Level Virtual Machines. IEEE Transactions on Dependable and Secure Computing. 10(2): 70-83 (2013)
[18] Zhiyong Shan, Xin Wang, Tzi-cker Chiueh: Enforcing Mandatory Access Control in Commodity OS to Disable Malware. IEEE Transactions on Dependable and Secure Computing 9(4): 541-555 (2012)
[19] Zhiyong Shan, Xin Wang, Tzi-cker Chiueh, Xiaofeng Meng: Facilitating inter-application interactions for OS-level virtualization. In Proceedings of the 8th ACM Annual International Conference on Virtual Execution Environments (VEE'12), 75-86
[20] Zhiyong Shan, Xin Wang, Tzi-cker Chiueh, and Xiaofeng Meng. "Safe side effects commitment for OS-level virtualization." In Proceedings of the 8th ACM international conference on Autonomic computing (ICAC'11), pp. 111-120. ACM, 2011.
[21] Zhiyong Shan, Xin Wang, and Tzi-cker Chiueh. 2011. Tracer: enforcing mandatory access control in commodity OS with the support of light-weight intrusion detection and tracing. In Proceedings of the 6th ACM Symposium on Information, Computer and Communications Security (ASIACCS '11). ACM, New York, NY, USA, 135-144. (full paper acceptance rate 16%)
[22] Shan, Zhiyong, Tzi-cker Chiueh, and Xin Wang. "Virtualizing system and ordinary services in Windows-based OS-level virtual machines." In Proceedings of the 2011 ACM Symposium on Applied Computing, pp. 579-583. ACM, 2011.
[23] Shan, Zhiyong, Yang Yu, and Tzi-cker Chiueh. "Confining windows inter-process communications for OS-level virtual machine." In Proceedings of the 1st EuroSys Workshop on Virtualization Technology for Dependable Systems, pp. 30-35. ACM, 2009.
[24] Shan, Zhiyong. "Compatible and Usable Mandatory Access Control for Good-enough OS Security." In Electronic Commerce and Security, 2009. ISECS'09. Second International Symposium on, vol. 1, pp. 246-250. IEEE, 2009.
[25] Xiao Li, Wenchang Shi, Zhaohui Liang, Bin Liang, Zhiyong Shan. Operating System Mechanisms for TPM-Based Lifetime Measurement of Process Integrity. Proceedings of the IEEE 6th International Conference on Mobile Adhoc and Sensor Systems (MASS 2009), Oct., 2009, Macau SAR, P.R.China, IEEE Computer Society. pp. 783–789.
[26] Xiao Li, Wenchang Shi, Zhaohui Liang, Bin Liang, Zhiyong Shan. Design of an Architecture for Process Runtime Integrity Measurement. Microelectronics & Computer, Vol.26, No.9, Sep 2009:183~186. (in Chinese)
[27] Zhiyong Shan, Wenchang Shi. "STBAC: A New Access Control Model for Operating System". Journal of Computer Research and Development, Vol.45, No.5, 2008: 758~764.(in Chinese)
[28] Liang Wang, Yuepeng Li, Zhiyong Shan, Xiaoping Yang. Dependency Graph based Intrusion Detection. National Computer Security Conference, 2008. (in Chinese)
[29] Zhiyong Shan, Wenchang Shi. "An Access Control Model for Enhancing Survivability". Computer Engineering and Applications, 2008.12. (in Chinese)
[30] Shi Wen Chang, Shan Zhi-Yong. "A Method for Studying Fine Grained Trust Chain on Operating System", Computer Science, Vol.35, No.9, 2008, 35(9):1-4. (in Chinese)
[31] Liang B, Liu H, Shi W, Shan Z. Automatic detection of integer sign vulnerabilities. In International Conference on Information and Automation, ICIA 2008. (pp. 1204-1209). IEEE.
[32] Zhiyong Shan, Qiuyue Wang, Xiaofeng Meng. "An OS Security Protection Model for Defeating Attacks from Network", the Third International Conference on Information Systems Security (ICISS 2007), 25-36.
[33] Zhiyong Shan, "A Security Administration Framework for Security OS Following CC", Computer Engineering, 2007.5, 33(09):151-163. (in Chinese)
[34] Shan Zhiyong, "Research on Framework for Multi-policy", Computer Engineering, 2007.5, 33(09):148-160. (in Chinese)
[35] Zhiyong Shan, Shi Wenchang, Liao Bin. "Research on the Hierarchical and Distributed Network Security Management System". Computer Engineering and Applications, 2007.3, 43(2):20-24. (in Chinese)
[36] Zhiyong Shan, "An Architecture for the Hierarchical and Distributed Network Security Management System", Computer Engineering and Designing, 2007.7, 28(14):3316-3320. (in Chinese)
[37] Shan Zhi Yong, Sun Yu Fang, "Study and Implementation of Double-Levels-Cache GFAC", Chinese Journal of Computers, Nov, 2004, 27(11):1576-1584. (in Chinese)
[38] Zhiyong Shan, Yufang Sun, "An Operating System Oriented RBAC Model and Its Implementation", Journal of Computer Research and Development, Feb, 2004, 41(2):287-298. (in Chinese)
[39] Zhiyong Shan, Yufang Sun, "A Study of Extending Generalized Framework for Access Control", Journal of Computer Research and Development, Feb, 2003, 40(2):235-244. (in Chinese)
[40] Shan Zhi Yong, Sun Yu Fang, "A Study of Generalized Environment-Adaptable Multi-Policies Supporting Framework", Journal of Computer Research and Development, Feb, 2003, 40(2):228-234. (in Chinese)
[41] Shan Zhiyong, Research on the Framework for Multi-Policies and Practice in Secure Operation System. Phd Thesis, Institute of Software, Chinese Academy of Science 2003. (in Chinese)
[42] Shan Zhi Yong, Sun Yu Fang, "A Study of Security Attributes Immediate Revocation in Secure OS", Journal of Computer Research and Development, Dec, 2002, 39(12):1681-1688. (in Chinese)

[43] Shi Wen Chang, Sun Yu Fang, Liang Hong Liang, Zhang Xiang Feng, Zhao Qing Song, Shan Zhi Yong. Design and Implementation of Secure Linux Kernel Security Functions. Journal of Computer Research and Development, 2001, Vol.38, No.10, 1255-1261.
[44] Zhiyong Shan, Tzi-cker Chiueh, Xin Wang. Duplication of Windows Services. CoRR, 2016.
[45] Zhiyong Shan. Suspicious-Taint-Based Access Control for Protecting OS from Network Attacks. Technical Report, 2014.
[46] Zhiyong Shan, Bin Liao. Design and Implementation of A Network Security Management System. Technical Report, 2014.
[47] Zhiyong Shan. A Study on Altering PostgreSQL From Multi-Processes Structure to Multi-Threads Structure. Technical Report, 2014.
[48] Zhiyong Shan. Implementing RBAC model in An Operating System Kernel. Technical Report, 2015. (in Chinese)
[49] Zhiyong Shan. A Hierarchical and Distributed System for Handling Urgent Security Events. Technical Report, 2014. (in Chinese)
[50] Zhiyong Shan. An Review On Thirty Years of Study On Secure Database and It`s Architectures. Technical Report, 2014. (in Chinese)
[51] Zhiyong Shan. An Review on Behavior-Based Malware Detection Technologies on Operating System. Technical Report, 2014.